\documentstyle[sprocl,epsfig]{article}
\def\be{\begin{equation}}
\def\ee{\end{equation}}
\def\bea{\begin{eqnarray}}
\def\eea{\end{eqnarray}}

\begin{document}
\begin{flushright}
YUMS - 98/05\\
February 1997\\
\end{flushright}
\title{IMPLICATIONS OF THE HERA EVENTS FOR THE R-PARITY BREAKING 
SUSY SIGNALS AT TEVTARON}
\author{Monoranjan Guchait} 
\address{Department of Physics, Yonsei University, Seoul 120-749, 
South Korea}
%%%%%%%%%%%%%%%%%%%%%%%%%%%%%%%%
%
% You may repeat \author \address
% as often as necessary
%
%%%%%%%%%%%%%%%%%%%%%%%%%%%%%%%
\maketitle
\abstracts{ I report here on the systematic analysis at Tevatron of 
the R-parity violating SUSY signals which correspond to the possible 
solution of anomalous HERA events}

%\section{Guidelines}
%\subsection{Producing the Hard Copy}
%\label{subsec:prod}
The two experimental groups at DESY, H1 and ZEUS, have reported some anomalous 
high $Q^2$ events from the $e^{+}p$ collider. 
The H1 has seen 12 
neutral current events, at $Q^2 >$ 15,000 $GeV^2$ against the 
Standard Model(SM)
prediction of 5 events~\cite{ad} ; 
while the ZEUS has reported 5 events at $Q^2 >$20,000 $GeV^2$ against the 
prediction~\cite{br} of 2. The excess of 7 events observed by H1 has 
clusterd around a common $e^{+}q$ mass of 
\be
M \simeq 200 GeV
\ee
which is inconsistent to ZEUS events~\cite{wo}.  
These are based on 1994-'96 data and corresponding to a combined luminosity
34$pb^{-1}$ for the two experiments. Now taking together, there are 10 excess
events which corresponds to a cross section 
\be
\sigma \simeq 0.4 pb 
\label{eq:cs}
\ee
for the 80\% detection
efficiency for each experiments.

Many mechanisms of new physics was proposed to explain these ~\cite{chou,dre,
alt,bel} new events.
Among the proposed mechanisms of new physics,  
the R-parity violating(RPV) SUSY model is the most popular one. 
The purpose of this work~\cite{guchait} is to study the implications of 
the RPV SUSY model at Tevatron 

The Minimal Supersymmetric Standard Model(MSSM) with explicit R-parity 
breaking~\cite{nille} contains the following Yukawa terms in the Lagrangian
\be
L = \lambda_{ijk} l_i \tilde l_j \bar e_k + \lambda'_{ijk} l_i \tilde
q_j \bar d_k + \lambda''_{ijk} \bar d_i \tilde{\bar d}_j \bar u_k ,
\ee
plus analogous terms from the permutation of the tilde, denoting the
scalar superpartner. We have used the same notation as of ref~\cite{guchait}. 
The terms relevant for the HERA
events are
\be
\lambda'_{1jk} (l_1 \tilde q_j \bar d_k + l_1 q_j \tilde{\bar d}_k) + h.c.
\ee
Therefore, for a given quark flux 
one can easily calculate the cross section for these process at HERA
when $\lambda' \sqrt B$ is given, where B denotes the squark branching 
fraction into the shown channel. While the size $\lambda' \sqrt B$ 
is constrained by 
HERA data as given by Eq.~\ref{eq:cs} and the value of $\lambda'$ 
is restricted  by its upper bounds 
from some other process~\cite{dre,guchait}. 
It is shown~\cite{chou,dre} that the viable couplings which are allowed by the 
HERA data and as well as by its corresponding 
existing upper limits 
are the $\lambda'_{121}$ and $\lambda'_{131}$ which stand for the scharm and
stop squark production from a valence quark i.e . 
\be
e ^+d \rightarrow \tilde c_L \rightarrow e ^+d.  
\label{eq:spa}
\ee
\be
e ^+d \rightarrow \tilde t_L \rightarrow e ^+d ,
\label{eq:murnf}
\ee
where the L subscript stands for left chirality. 
Therefore, we have concentrated only on these
two scenarios and called them scharm and stop scenario for corresponding 
to the $\lambda'_{121}$ and $\lambda'_{131}$ couplings respectively.

Since the product of $\lambda' \sqrt B$ is constrained 
by the cross section, Eq.~\ref{eq:cs},
of the HERA events, it is possible to express
B in terms of the MSSM parameters under the assumption that there is 
only one dominant RPV coupling~\cite{guchait}. 
The dependence of B on the MSSM parameter space for a given scenario
will be discussed below.

It is to be noted that in our analysis we assume a common gaugino and a 
common sfermion mass at the unification scale which results the masses 
of gauginos ($\tilde g,\tilde W$ and $\tilde B$) are related via their gauge
couplings at the electroweak unification scale~\cite{guchait}.
Under this assumption we have investigated the possibility of finding 
the stop and scharm scenario
at the Tevatron collider. The results are in order.

\subsection{The Stop Scenario at Tevatron:}
The dominant production mechanism for stop at Tevatron 
are the leading order QCD process of quark-antiquark and gluon-gluon
fusion~\cite{kane}
\be
\bar q q \rightarrow \tilde{\bar t}_1 \tilde t_1 \ \ , \ \ gg
\rightarrow \tilde{\bar t}_1 \tilde t_1 .
\ee
Because of the negligble contribution of the RPV couplings, we have not
considered it here.

The dominant decay mode of stop is the RPV mode 
\be
\tilde t_1 \rightarrow e^+ d .
\label{eq:stdr}
\ee
as long as $M_{\tilde t_1} \simeq M_{\tilde W_1}$, because the decay 
mode in the 
neutralino channel is higher order process due to the large top 
quark mass. But the R-parity conserving decay mode  
\be
\tilde t_1 \rightarrow b \tilde W_i
\ee
will dominate when it will be kinematically allowed.
\begin{figure}[htb]
%\vspace*{3.1cm}
\centerline{\epsfig{file=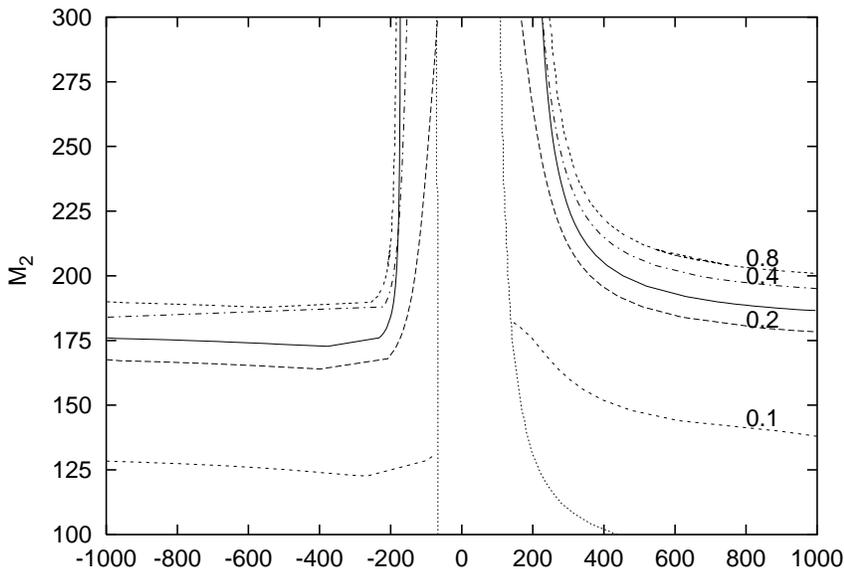,height=8cm}}
%\vspace{-0.5cm}
\caption
{The stop branching fraction (B) is shown as a contour plot in
the $M_2$, $\mu$ plane for the stop mass $M_{\tilde t_1}= 200 GeV$,
$tan\beta = 2$ and $\theta_ {\tilde t} = 0^\circ$. 
Contours for the lighter chargino mass of 180GeV(solid line)
and 85GeV(dotted line) are shown.}
\end{figure}
%\begin{minipage}{5cm}
%\epsfig{file=fig1a.ps,height=4cm}
%\end{minipage}
%\begin{minipage}{5cm}
%\epsfig{file=fig1a.ps,height=4cm}
%\end{minipage}
%\end{figure}
In Fig.1, the values of B are shown over the relevent MSSM parameter 
space. It is clear from the fig.1 that the the signal of stop at
Tevatron depends very crucially on the MSSM parameter space.
In our analysis, we have considered three types of decay configuration 
of stop which are discussed below.

(a) Direct Leptonic decay: In this case both the stop squarks decay
in the RPV channel(Eq.~\ref{eq:stdr}) resulting a pair of hard and 
isolated dilepton along with a pair of jets. 
The corresponding 95\% C.L. limits of 3 events  as obtained from the 
1st generation of Leptoquark search gives a mass bound by CDF
M$>$ 210 GeV~\cite{cdf}
where as this value for the D0 group~\cite{d0} is M$> 225 GeV$. Thus
CDF and D0 leptoquark limits would rule out the stop scenario if the
direct leptonic decays are dominant, i.e. $M_{\tilde W_1} \ge   
M_{\tilde t_1}$. However, dilepton cross section 
for the B$\le 0.2$ is so small that it is not possbile to probe them in the
present data set of CDF and D0. But in the MI run one can probe the signal
cross section down to B = 0.2. Nonethless, there is a significant MSSM 
parameter space where B$\le 0.2$ and it is possible to probe this 
parameter space 
in the MI run. But it may be possible to probe this entire MSSM 
parameter space
at TEV33 where luminosity will be 20 $fb^{-1}$.

(b)Mixed Mode:This mode corresponds to direct decay of one stop
(eq.~\ref{eq:stdr}) 
and other undergoes cascade decay as
\be
\tilde t_1 \rightarrow b \tilde W_1, \tilde W_1 \rightarrow q' \bar q
\tilde Z_1 , \tilde Z_1 \buildrel{\lambda'_{131}}\over{\rightarrow}
\bar b \bar\nu d + h.c.
\label{eq:tcas}
\ee
Thus the final states consists of a very hard $e^{\pm}$, a large number 
of jets including a pair of b and a modest amount of missing $p_{T}$
carried by the neutrino. The dominant background to this process is from 
$t \bar t$. Fig.2 shows the signal cross section for a stop mass 200 GeV
along with the $t \bar t$ background at $\sqrt s = 2 TeV$ for a following
MSSM parameters,
\be
M_2 = 150 GeV, \ \mu = -400 GeV, \ \tan\beta = 2 \Rightarrow M_{\tilde
W_1} = 158 GeV, \ M_{\tilde Z_1} = 77 GeV;
\ee
\begin{figure}[htb]
\vspace{-1.8cm}
\centerline{\epsfig{file=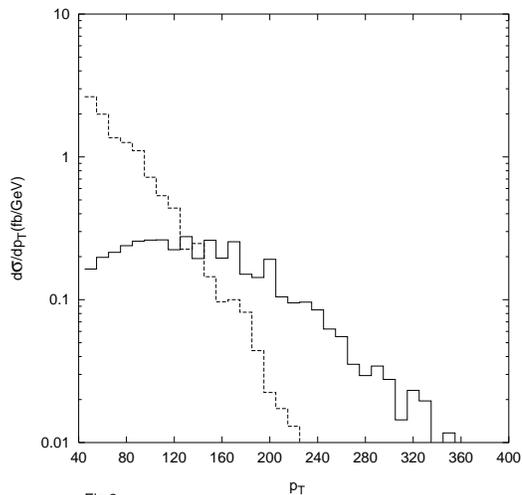,height=11cm}}
\vspace{-1.9cm}
\caption{
The signal cross section(solid line) corresponding to the mixed
mode(b) is shown for stop mass of 200 GeV along with the $t \bar t$
background(dashed line)
against the $p_T$ of the electron at 2 TeV. The MSSM parameters are $M_2 =
150 GeV$,
$\mu = -400 GeV$, and $tan\beta = 2$. }
\end{figure}
Note that this signal rate is quite insensitive of MSSM parameters. From  
Fig.2 it is clear that the signal can be separated from the background 
by exploiting the clustering of invariant mass of the electron
with the hardest jets at $M\simeq 200 GeV$. The size of the signal in Fig.2
is about 50fb which means few events are expecetd for the MI run for the 
MSSM parameter space ( in Fig.1) where the branching 
fraction is B$\simeq 7\%$.  

(c)Cascade decays: It corresponds to the cascade decays, eq.~\ref{eq:tcas} ,of 
both the stop squarks resulting in the final states four b quarks and a missing
$p_T$ carried by the neutrinos. We need to do the tripple b tagging to 
suppress background which is mainly 
from $t\bar t$ and $b\bar b$ production.
We have estimated the  
signal cross section for two different channels (i) 3 b quarks + missing 
$p_T$ and (ii) 1 lepton + 3 b + missing $p_T$.  
We found no viable signature for detecting the stop pair production
for the integrated luminosity 2 $fb^{-1}$. 
But in the MI run it may be possible to probe entire MSSM parameter space
for larger b-tagging efficiency.

\subsection{The Charm Squark( Scharm!) Scenario at Tevatron}
In this case one expects 8 species of roughly degenerate squarks
along with a gluino of comaparable or smaller mass. Only one of them,
the left handed charm squark $\tilde c_L$, has the R violating
decay mode 
\be
\tilde c_L \rightarrow e^+ d .
\label{eq:crs}
\ee
as required to explain the HERA anomaly. However, scharm has also
R conserving decay modes which are  
\be
\tilde c_L \rightarrow s \tilde W_i, s \tilde Z_i
\ee
We have seen that the corresponding branching fraction B for the decay
modes eq.~\ref{eq:crs} is B$\le 1/20$. Therefore, the direct decay 
channel eq.~\ref{eq:crs} will not give any viable SUSY signal.
We have considered the cascade decays of the squarks and gluinos 
into the LSP which gives one electron through its decay as
\be
\tilde Z_1 \buildrel{\lambda'_{121}}\over{\rightarrow} \bar c e^+ d
(\bar s \bar \nu d) + h.c.
\label{eq:lspd}
\ee
As a consequence, the pair of LSP provides like sign dilepton(LSD)
signature in this case due to its Majorana nature. 
We have considered the all possible strong production 
process~\cite{kane} like pair of gluino, pair of squark and associated 
production~\cite{guchait} and their subsequent decays according to 
the masses of electroweak gauginos~\cite{guchait}.
Considering only the $\lambda'_{121}$ as leading RPV coupling, as suggested 
by the scharm scenario, we have estimated the LSD rates for a given value
of charm squark mass(=210 GeV) applying some kinematic cuts 
to suppress the SM background.
We have seen that one can expects to see at least half of dozen
LSD events with the present Tevatron luminosity of 110 $pb^{-1}$. Finally,
we conclude with a expectation that the Tevatron group
will analyse their data to probe the RPV SUSY model, and in particular
to test the charm squark scenario for the anomalous HERA events. 
 
The work reported here is done with collaboration of 
D.P.Roy~\cite{guchait}.

\end{document}